\documentclass[conference]{IEEEtran}

\begin{document}
\input{epsf}

\title{Single-Scan Min-Sum Algorithms \\for Fast Decoding of LDPC Codes}

\author{\authorblockN{Xiaofei Huang} \\
\authorblockA{School of Information Science and Technology\\Tsinghua University, Beijing, P.~R.~China, 100084 \\
Email: huangxiaofei@ieee.org\\
(Accepted by IEEE Information Theory Workshop, Chengdu, China, 2006)}
}
%

\maketitle

\begin{abstract}
Many implementations for decoding LDPC codes
	are based on the (normalized/offset) min-sum algorithm 
	due to its satisfactory performance and simplicity in operations.
Usually, each iteration of the min-sum algorithm contains two scans, the horizontal scan and the vertical scan.
This paper presents a single-scan version of the min-sum algorithm to speed up the decoding process.
It can also reduce memory usage or wiring 
	because it only needs the addressing from check nodes to variable nodes
	while the original min-sum algorithm requires that addressing plus the addressing from variable nodes to check nodes. 
To cut down memory usage or wiring further,
	another version of the single-scan min-sum algorithm is presented
	where the messages of the algorithm are represented by single bit values 
	instead of using fixed point ones.
The software implementation has shown 
	that the single-scan min-sum algorithm is more than twice as fast as the original min-sum algorithm.
\end{abstract}

\section{Introduction}
	

The sum-product algorithm~\cite{Kschischang01,Aji00}, also known as the belief propagation algorithm~\cite{Pearl88}, 
	is the most powerful iterative soft decoding algorithm for LDPC (low density parity check) codes~\cite{Gallager:LDPCC:thesis,MacKay:GCBOVSM,Richardson:DOCAILDPCC}.
The normalized/offset min-sum algorithm~\cite{Wiberg:thesis,Fossorier99,Jinghu2002,JinghuChenThesis} 
	has demonstrated in \cite{Jinghu2002,JinghuChenThesis} as a good approximation to the sum-product algorithm.
It is a parallel, iterative soft decoding algorithm for LDPC codes.
It is simpler in computation than the sum-product algorithm
	because it uses only minimization and summation operations 
	instead of multiplication and summation operations used by the latter.
It is also simpler in computation than the sum-product algorithm in the log domain 
	because the latter uses non-linear functions.
For hardware/software implementations, multiplication operations
	and non-linear functions are, in general, more expensive than minimization and summation operations.
	
Despite of its reduced complexity,
	we found out, in implementing the normalized/offset min-sum algorithm for China's HDTV, 
	that the min-sum algorithm is still expensive for hardware/software implementations
	since two scans are required by the algorithm at each iteration, 
	and its convergence rate is generally not satisfactory.
The min-sum algorithm is also not memory efficient.
The temporary results of the algorithm are stored in memory as fixed point values.
The number of values is proportional to 
	the number of non-zero elements of the parity check matrix of a LDPC code.
They require large circuit areas because the number of nonzero elements is not small in practice.
Manipulating those values also takes a lot of system time and consumes much of system power at run-time.
We concluded that further simplification of the min-sum algorithm is needed 
	to suit the ever demanding requirements of the next generation communication systems.

This paper presents two simplified versions of the min-sum algorithm
	to increase its decoding speed and reduce the requirement on memory usage.
Those simplifications are based on several obvious observations 
	with some of them already mentioned by other researcher~\cite{GuilloudThesis}.
However, no detail has been offered in the previous literature in the form of algorithms
	which can be directly used by engineers and practitioners in the communication area.
Furthermore, the advantage of the simplifications is not neglectable
	because the simplified min-sum algorithm more than doubles the decoding speed
	of the standard min-sum algorithm in our software implementation (in C language)
	for decoding the quasi-cyclic irregular LDPC codes used for China's HDTV,
	the irregular LDPC codes used for European digital video broadcasting using satellites (DVB-S2), 
	and the regular/irregular LDPC codes from Dr.~MacKay's website.
The comparison is fair because both algorithms are simple in operations 
	and can be implemented in software in a straightforward way 
	without much room for further improvement.
Our HDTV research group at Tsinghua University has benefited from the simplifications because 
	we most often use software simulations first to test the performance of different LDPC codes for China's HDTV
	which could be very time-consuming and may take hours or even days running on fast Intel-based desktop computers.
	
\section{Definitions and notations}
LDPC codes belong to a special class of linear block codes
	whose parity check matrix $H$ has a low density of ones.
LDPC codes were originally introduced by Gallager in his thesis~\cite{Gallager:LDPCC:thesis}.
After the discovery of turbo codes in 1993 by Berrou et al.~\cite{Berrou93},
	LDPC codes were rediscovered by Mackay and Neal~\cite{MacKay:GCBOVSM} in 1995.
Both classes have excellent performances in terms of error correction close to the Shannon limit.
%
For a binary LDPC code, $H$ is a binary matrix with elements, denoted as $h_{mn}$ in $\{0,1\}$.
Let the code word length be $N$,
then $H$ is a $M \times N$ matrix, where $M$ is the number of rows. 
Each row $H_i$ ($1 \le i \le M$) of $H$ introduces one parity check constraint on input data $x=(x_1, x_2, \ldots, x_n)$,
	i.e., 
\[ H_i x^T = 0~mod~2 . \]
So there are $M$ constraints on $x$ in total.

Let ${\cal N}(m)$
be the set of variable nodes that are included in the $m$-th parity check constraint.
Let ${\cal M}(n)$
	be the set of check nodes which contain the variable node $n$.
${\cal N}(m) \setminus n$ denotes the set of variable nodes excluding node $n$ 
	that are included in the $m$-th parity check constraint.
${\cal M}(n) \setminus m$ stands for the set of check nodes excluding the check node $m$ 
	which contain the variable node $n$.
The symbol `$\setminus$' denotes the set minus.

For an additive white Gaussian noise channel and a binary modulation,
	let $y_n$ be the received data bit at position $n$,
\[ y_n = (-1)^{x_n} + \xi_n \ , \]
where $\xi_n$ is the channel noise.
The initial Log-likelihood ratio (LLR) for the input data bit $n$, denoted as $Z^{(0)}_n$, is 
\[ Z^{(0)}_n \equiv \ln \frac{p(x_n = 0/y_n)}{p(x_n = 1/y_n)}= 2y_n /\sigma^2 \ , \]
where $\sigma^2$ is the estimated variance of the channel noise.
The performance of the min-sum algorithm does not depend on the channel estimate.
We can, thus, set $Z^{(0)}_n = y_n$ in practice.

\section{The (Normalized/Offset) Min-sum Algorithm}

The standard min-sum algorithm~\cite{Wiberg:thesis,Fossorier99} for decoding LDPC 
	is a parallel, iterative soft decoding algorithm.
At each iteration, messages first are sent from the variable nodes to the check nodes,
	called the horizontal scan.
Then messages are sent from the check nodes back to the variable nodes, 
	called the vertical scan.
During each iteration, 
	the a-posteriori probability for each bit is also computed.
A hard decoding decision is made for each bit based on the probability,
	and decoded bits are checked against all parity check constraints to see if they are a valid codeword.

At iteration $k$, let $Z^{(k)}_n$ be the posteriori LLR for the input data bit $n$.
Let $Z^{(k)}_{mn}$ denote the message sent from variable node $n$ to check node $m$.
$Z^{(k)}_{mn}$ is the log-likelihood ratio
	that the $n$-th bit of the input data $x$ has the value $0$ versus $1$, 
	given the information obtained via the check nodes other than check node $m$.
Let $L^{(k)}_{mn}$ denote the message sent from check node $m$ to variable node $n$.
$L^{(k)}_{mn}$ is the log-likelihood ratio that the check node $m$ is satisfied 
	when the input data bit $n$ is fixed to value $0$ versus value $1$
	and the other bits are independent with log-likelihood ratios $Z_{mn^{'}}, n^{'} \in {\cal N}(m) \setminus n$.
The pseudo-code for the min-sum algorithm is given as follows.

{\noindent {\em Initialization :}}
For $n \in \{1, 2, \ldots, N\}$,
\[ Z^{(0)}_{mn} = Z^{(0)}_n, \quad \mbox{ for $m \in {\cal M}(n)$}. \]
{\em Iteration} 
\begin{enumerate}
\item {\em Horizontal scan (check node update rule) :}

For each $m$ and each $n \in {\cal N}(m)$, 
\begin{equation}
L^{(k)}_{mn} = \prod_{n^{'} \in {\cal N}(m) \setminus n} \mbox{sgn}(Z^{(k-1)}_{mn^{'}}) \cdot \min_{n^{'} \in {\cal N}(m) \setminus n} |Z^{(k-1)}_{mn^{'}}|
\label{MS_horizontal}
\end{equation}

\item {\em Vertical scan (variable node update rule) :}  

\begin{equation}
Z^{(k)}_{mn} = Z^{(0)}_n + \sum_{m^{'} \in {\cal M}(n) \setminus m} L^{(k)}_{m^{'}n} \ . 
\label{MS_vertical}
\end{equation}

\item {\em Decoding :}

For each bit, compute its posteriori log-likelihood ratio (LLR)
\[ Z^{(k)}_n = Z^{(0)}_n + \sum_{m^{'} \in {\cal M}(n)} L^{(k)}_{m^{'}n} \ . \]
Then estimate the original codeword ${\hat x}^{(k)}$ as
\[
{\hat x}^{(k)}_n = \left\{ \begin{array}{ll}
              0, & \mbox{if $Z^{(k)}_n > 0$}; \\
              1, & \mbox{otherwise};
              \end{array}
      \right. \quad \mbox{for $n = 1, 2, \ldots, N$} \ .
\]
If $H~({\hat x}^{(k)})^T = 0$ or the iteration number exceeds some cap, 
	stop the iteration and output ${\hat x}^{(k)}$ as the decoded codeword.
\end{enumerate}

One performance improvement to the above standard min-sum algorithm
	is to multiply $L^{(k)}_{mn}$ computed in (\ref{MS_horizontal}) by a positive constant $\lambda_k$ smaller than $1$,
	i.e.,
\[ L^{(k)}_{mn} \Leftarrow  \lambda_k L^{(k)}_{mn} \ , \]
The min-sum algorithm with such a modification is referred to as the normalized min-sum algorithm~\cite{Jinghu2002,JinghuChenThesis}.

Another improvement to the standard min-sum algorithm,
	is to reduce the reliability values $L^{(k)}_{mn}$ computed in (\ref{MS_horizontal})
	by a positive value $\beta_k$, i.e.,
\[ L^{(k)}_{mn} \Leftarrow  \max \left(L^{(k)}_{mn} - \beta_k, 0 \right) \ , \]
The min-sum algorithm with such a modification is referred to as the offset min-sum algorithm~\cite{JinghuChenThesis}.
The difference between the standard min-sum algorithm and the normalized/offset one
	is minor for software/hardware implementations.

\section{The Single-Scan Min-Sum Algorithm}
It is very straightforward to rewrite the variable node update rule~(\ref{MS_vertical}) as
\[ Z^{(k)}_{mn} = Z^{(k)}_n - L^{(k)}_{mn} \ . \]
If we have computed $Z^{(k)}_n$, 
	the variable node message $Z^{(k)}_{mn}$ can be obtained from the check node message $L^{(k)}_{mn}$.
Hence, we can merge the horizontal scan and the vertical scan
	into a single horizontal scan where only the check node messages $L^{(k)}_{mn}$ are computed directly
	from $Z^{(k-1)}_{n}$ and $L^{(k-1)}_{mn}$.
In summary, the single-scan min-sum algorithm consists of the following major steps.

{\noindent {\em Initialization :}} $\quad L^{(0)}_{mn} = 0$. 

{\noindent {\em Horizontal scan (check node update rule) :}} 
\begin{eqnarray}
Z^{(k)}_n &= & Z^{(0)}_n \ , \nonumber \\
L^{(k)}_{mn} &= &\prod_{n^{'} \in {\cal N}(m) \setminus n} \mbox{sgn}(Z^{(k-1)}_n - L^{(k-1)}_{mn^{'}}) \times  \nonumber \\
	& &~~~\min_{n^{'} \in {\cal N}(m) \setminus n} |Z^{(k-1)}_n - L^{(k-1)}_{mn^{'}}|  \ , \label{MS_horizontal_ss} \\
Z^{(k)}_n & += & L^{(k)}_{mn} \ . \nonumber
\end{eqnarray}

{\noindent {\em Decoding :}} ${\hat x}^{(k)}_n = 0,~\mbox{if $Z^{(k)}_n > 0$};~{\hat x}^{(k)}_n = 1,~\mbox{otherwise}.$

Compared with the original double-scan min-sum algorithm, 
	the single-scan version could not only be possibly faster, but also be more memory efficient.
We can save memory by storing $Z^{(k)}_{n}$s, which are of $N$ items, instead of $Z^{(k)}_{mn}$, 
	which are of $N \cdot d_v$ (average variable node degree) items.
For software implementations, 
	the single-scan version
	needs only to store the addressing (indexing) from check nodes to variable nodes.
However, for the original version,
	both the addressing from check nodes to variable nodes and the one from variable nodes to check nodes are required.
The single-scan version cuts down the amount of memory used for addressing by half.
Such memory saving could be important if the min-sum algorithm is implemented 
	in next-generation wireless/mobile computing devices  
	where available memory could be very limited.

Could the memory saving be directly translated into the saving of wiring for hardware implementations?
It has been found that our hardware implementation 
	of the original min-sum algorithm for decoding the LDPC codes used for China's HDTV
	takes a significant amount of circuit area (sometimes $50\%$) 
	just for implementing the connections from the variable nodes to the check nodes 
	and the connections from the check nodes to variable nodes.
Since the simplified min-sum algorithm has only the horizontal scan,
	the circuit area could be reduced if only the connections from the check nodes to variable nodes are required.
We are now at the stage of verifying this statement in our lab.

\section{Further Simplification for the Single-Scan Algorithm}

The original min-sum algorithm
	uses a lot of memory for storing the variable node messages $Z^{(k)}_{mn}$ and the check node messages $L^{(k)}_{mn}$.
Although we can use one memory cell to store both $Z^{(k)}_{mn}$ and $L^{(k)}_{mn}$,
	we still need $\sum_n |{\cal M}(n)|$ memory cells to store them,
	one for each non-zero element of the parity check matrix $H$.
This statement still holds for the single-scan min-sum algorithm 
	where only the check node messages $L^{(k)}_{mn}$ are stored.
	
If we use $b$ bits ($b = 6 \sim 8$ in practice) to store a value (containing its sign), 
	then in total they require $b \cdot \sum_n |{\cal M}(n)|$ bits.
For VLSI implementations, that could take a significant amount of circuit area.
It also leads to high energy consumption due to the intensive manipulation (reading/writing) 
	of those memory cells at each iteration, two writing operations for each cell by the original min-sum algorithm.

Our simplification comes from the following observation 
	of the check node messages $L^{(k)}_{mn}$ computed at the horizontal scan.
The check node messages $L^{(k)}_{mn}$
	is computed using (\ref{MS_horizontal}) or (\ref{MS_horizontal_ss}).
For the check node messages $L^{(k)}_{mn}$ of the same check node $m$, 
	all have the same absolute value except one.
The first absolute value is the minimal absolute value of $|L^{(k-1)}_{mn}|$s, for the same $m$.
The second absolute value is the second minimal absolute value of $|L^{(k-1)}_{mn}|$s.
This observation is quite obvious.
Dr.~Guilloud also mentioned in his thesis~\cite{GuilloudThesis} (section 4.1.2) 
	that two messages need to be saved for each parity-check equation.
This paper offers the detail of exploring this unique characteristic 
	to possibly cut down the memory usage of the single-scan min-sum algorithm further.
	
To be more specific, for the check node messages $L^{(k)}_{mn}$ of the same check node $m$, 
	let $A^{(k)}_m$ be the first absolute value,
\[ A^{(k)}_m \equiv \min_{n \in {\cal N}(m)} |L^{(k)}_{mn}| \ . \]
From Eq.~(\ref{MS_horizontal}), we have
\[ A^{(k)}_m = \min_{n \in {\cal N}(m)} |Z^{(k-1)}_{mn}| \ , \]
which is the minimal value of $|Z^{(k-1)}_{mn}|$s of the same check node $m$.

Let $B^{(k)}_m$ be the second absolute value,
\[ B^{(k)}_m \equiv \max_{n \in {\cal N}(m)} |L^{(k)}_{mn}| \ , \]
and let ${\tilde n}^{(k)}_m$ be the position of $L^{(k)}_{mn}$ of the second minimal absolute value, 
\[ {\tilde n}^{(k)}_m \equiv \arg \max_{n \in {\cal N}(m)} |L^{(k)}_{mn}| \ . \]

From Eq.~(\ref{MS_horizontal}), we have
\[ {\tilde n}^{(k)}_m = \arg \min_{n \in {\cal N}(m)} |Z^{(k-1)}_{mn}| \ . \]
That is, the position of $L^{(k)}_{mn}$ of the second minimal absolute value $B^{(k)}_m$, ${\tilde n}^{(k)}_m$,
	is at the position of $Z^{(k-1)}_{mn}$ of the minimal absolute value.

From Eq.~(\ref{MS_horizontal}), we also have
\[ B^{(k)}_m = \min_{n \in {\cal N}(m) \setminus {\tilde n}^{(k)}_m} |Z^{(k-1)}_{mn}| \ , \]
	which is the second minimal value of $|Z^{(k)}_{mn}|$s of the same check node $m$.

Hence, for the check node messages $L^{(k)}_{mn}$ of the same check node,
	to save memory, we only need to store the two absolute values, the position of the first one, 
	and the signs of $L^{(k)}_{mn}$.

Let the sign of $L^{(k)}_{mn}$ be $s^{(k)}_{mn}$, $s^{(k)}_{mn} = \mbox{sgn} (L^{(k)}_{mn})$.
Let $f(A, B, n, {\tilde n})$ be a function defined as
\[
f(A, B, n, {\tilde n}) = \left\{ \begin{array}{ll}
                 B, & \mbox{if $n = {\tilde n}$}; \\
                 A, & \mbox{otherwise}.
                 \end{array}
         \right. 	
\]
The check node message $L^{(k)}_{mn}$ can be recovered from its sign $s^{(k)}_{mn}$, 
	the two absolute values, $A^{(k)}_m$ and $B^{(k)}_m$,
	and the position ${\tilde n}^{(k)}_m$ as
\[ L^{(k)}_{mn} = s^{(k)}_{mn} f(A^{(k)}_m, B^{(k)}_m, n, {\tilde n}^{(k)}_m) \ . \]

In the single-scan min-sum algorithm presented in the previous section,
	substituting the computation of $L^{(k)}_{mn}$ 
	by the computation of $s^{(k)}_{mn}$, $A^{(k)}_m$, $B^{(k)}_m$, and ${\tilde n}^{(k)}_m$,
	we have a memory efficient version of the single-scan min-sum algorithm.

{\noindent {\em Initialization :}} 
$A^{(0)}_m  = B^{(0)}_m = {\tilde n}^{(0)}_m = 0,~~s^{(0)}_{mn} = \mbox{sgn}(L^{(0)}_n)$. 

{\noindent {\em Horizontal scan (check node update rule) :}} 
\begin{eqnarray}
Z^{(k)}_n &= & Z^{(0)}_n \ , \nonumber \\
L^{(k-1)}_{mn} & = & s^{(k-1)}_{mn} f(A^{(k-1)}_m, B^{(k-1)}_m, n, {\tilde n}^{(k-1)}_m) \ , \nonumber \\
L^{(k)}_{mn} &= &\prod_{n^{'} \in {\cal N}(m) \setminus n} \mbox{sgn}(Z^{(k-1)}_n - L^{(k-1)}_{mn^{'}}) \times  \nonumber \\
	& &~~~\min_{n^{'} \in {\cal N}(m) \setminus n} |Z^{(k-1)}_n - L^{(k-1)}_{mn^{'}}|  \ , \nonumber \\
Z^{(k)}_n & += & L^{(k)}_{mn} \ , \nonumber \\
A^{(k)}_m & = & \min_{n \in {\cal N}(m)} |L^{(k)}_{mn}| \ , \label{one_bit_computing_e1} \\
B^{(k)}_m &=& \max_{n \in {\cal N}(m)} |L^{(k)}_{mn}| \ , \label{one_bit_computing_e2} \\
{\tilde n}^{(k)}_m &=& \arg \max_{n \in {\cal N}(m)} |L^{(k)}_{mn}| \ , \nonumber \\
s^{(k)}_{mn} &=& \mbox{sgn} (L^{(k)}_{mn}) \ . \nonumber 
\end{eqnarray}

{\noindent {\em Decoding :}} ${\hat x}^{(k)}_n = 0,~\mbox{if $Z^{(k)}_n > 0$};~{\hat x}^{(k)}_n = 1,~\mbox{otherwise}. $

In the memory efficient version of the single-scan min-sum algorithm, 
	the previous check node messages $L^{(k-1)}_{mn}$ of the check node $m$ 
	are computed on fly and stored as temporary data during the computation of 
	check node messages $L^{(k)}_{mn}$ of the same check node $m$.
$s^{(k)}_{mn}$, $A^{(k)}_m$,$B^{(k)}_m$, ${\tilde n}^{(k)}_m$, and $Z^{(k)}_n$ are persistent data stored in memory.

The single-scan min-sum algorithm is fully equivalent to the original min-sum algorithm.
To modify the single-scan min-sum algorithm to be equivalent to the normalized min-sum algorithm,
	Eq.~(\ref{one_bit_computing_e1}) and Eq.~(\ref{one_bit_computing_e2}) should be changed to
\begin{eqnarray*}
A^{(k)}_m & = & \lambda_k\min_{{n \in \cal N}(m)} |L^{(k)}_{mn}| , \label{one_bit_computing_e1_modified} \\
B^{(k)}_m & = & \lambda_k\max_{{n \in \cal N}(m)} |L^{(k)}_{mn}| . 
\end{eqnarray*}
where $\lambda_k$ is a constant at iteration $k$ satisfying $0 < \lambda_k < 1$.

To modify the single-scan min-sum algorithm to be equivalent to the offset min-sum algorithm,
	Eq.~(\ref{one_bit_computing_e1}) and Eq.~(\ref{one_bit_computing_e2}) should be changed to
\begin{eqnarray*}
A^{(k)}_m & = & \max(\min_{{n \in \cal N}(m)} |L^{(k)}_{mn}| - \beta, 0) , \label{one_bit_computing_e1_offset} \\
B^{(k)}_m & = & \max(\max_{{n \in \cal N}(m)} |L^{(k)}_{mn}| - \beta, 0) \ , 
\end{eqnarray*}
where $\beta$ is a constant, satisfying $\beta > 0$.

\section{Summary}

This paper presents the single-scan min-sum algorithm 
	as a simplified version of the original (normalized/offset) min-sum algorithm for decoding LDPC codes.
It merges the horizontal scan and the vertical scan in the original min-sum algorithm
	into a single horizontal scan.
A memory efficient version of the single-scan min-sum algorithm is also presented
	where the check node messages of each check node are stored using their signs together with two of the messages
	of the minimal absolute values.
All the simplifications are applicable for decoding binary LDPC codes.

\nocite{HuangISIT05,Yedidia05,Aji97,Pearl88,Kschischang01,Boutillon2000,Mansour2002,Howland2001,Yeo01high,Blanksby2002,Kim2002,Hocevar2003,ChenHocevar2003}
\bibliographystyle{../bib/IEEEtran}

\end{document}